\documentclass[final,5p,times,twocolumn]{elsarticle}

\usepackage{amssymb,amsmath,amsfonts}
\usepackage{graphicx,psfrag,color}
\usepackage{dcolumn}
\usepackage{bm}
\usepackage{mathbbol}

\journal{Physics Letters A}

\begin{document}

\begin{frontmatter}

\title{Almost perfect transport of an entangled two-qubit state through a spin chain}

\author{Rafael Vieira}
\author{Gustavo Rigolin}
\ead{rigolin@ufscar.br}
\address{Departamento de F\'isica, Universidade Federal de
S\~ao Carlos, 13565-905, S\~ao Carlos, S\~ao Paulo, Brazil}


\begin{abstract}
We show that using a slightly modified XX model for a spin-1/2 chain, one can transmit
almost perfectly a maximally entangled two-qubit state from one end of the chain to
the other one. This is accomplished without external fields or modulation of the coupling constants among the qubits.
We also show that this strategy works for any size of the chain and is relatively robust to imperfections
in the coupling constants among the qubits belonging to the chain. Actually, under certain scenarios of small disorder, 
we obtain better results than those predicted by the optimal ordered and noiseless case.
\end{abstract}



\end{frontmatter}


\section{Introduction} 

High fidelity transmissions of quantum states from one place to another are important ingredients needed
in the implementation of several quantum information tasks \cite{ben00}.  
Indeed, quantum communication protocols, and in particular quantum key distribution 
protocols \cite{ben84}, cannot work without a reliable transmission of a quantum state from one place
(Alice) to another (Bob). Even a to-be-built quantum computer will not work without a 
high fidelity quantum state transfer protocol within its hardware, since quantum information must flow
without much distortion among the many components of a quantum chip. 

There are at least three ways by which quantum information, here a synonym to a quantum state,
can be transmitted from Alice to Bob. The first one is the obvious direct transmission of the quantum state, where
the original physical system (a qubit, for simplicity) encoding the quantum information is sent from Alice to Bob. 
This usually happens when the quantum information is encoded in the state of a photon, which is sent along an optical fiber from Alice to Bob.
The second way to transmit quantum information is via the quantum teleportation protocol \cite{ben93}, where a highly entangled state
shared between Alice and Bob is the channel through which one is able to make a qubit with Bob be
described by the state originally describing Alice's qubit. 
A third possibility is to use spin chains connecting Alice and Bob as the channel through which the quantum state
describing one end of the chain ends up after a certain time describing the other end of it \cite{bos03}.
This is achieved by engineering the coupling constants among the qubits such that at time $t>0$ 
Bob's end of the chain is described by the state initially describing  Alice's end at time $t=0$.  
In this last strategy, as well as in the quantum teleportation protocol, the physical system
originally encoding the information is not sent from Alice to Bob, only the quantum state (quantum information) 
moves from Alice to Bob.

A main advantage of using the last strategy is related to the fact that once the coupling constants among the 
spins of the chain are set up to achieve a high fidelity transmission, we do not need to 
change them or switch them on and off. Such fixed arrangements can be very practical 
to allow the transmission of quantum states among the several components of a quantum computer, where 
it is not an easy task to constantly adjust the interaction strength among its qubits \cite{bos03}.
In addition to that, if the quantum chip is manufactured on a solid state system, it will be an advantage to have the 
communication channels connecting the several logic gates of the chip built on the same physical
system. In this way there will be no need to sophisticated interfacing between different physical systems as it happens,
for example, when one uses photons to transmit the information and spins to process it \cite{bos03}.

So far the great majority of works dealing with quantum state transmission have either studied
\begin{itemize}
\item [(a)] the transferring of a single excitation or an arbitrary qubit from Alice to Bob 
\cite{bos03,chr04,nik04,sub04,osb04,chr05,woj05,kar05,har06,huo08,gua08,ban10,kur11,apo12,lor13,hor14,shi15,zha16,che16,est17}; 
\item [(b)] the creation of a highly entangled state between Alice and Bob (the two ends or two specific sites of the chain) 
\cite{bos03,chr05,li05,har06,ban10,lor13,est17b}; 
\item [(c)] or the transferring of two (or more) excitations or two (or many)-qubit states from Alice to Bob 
\cite{nik04,sub04,chr05,shi05,lor13,sou14,lor15}.
\end{itemize}
%

\begin{figure}[!ht] \begin{center}
\includegraphics[width=8cm]{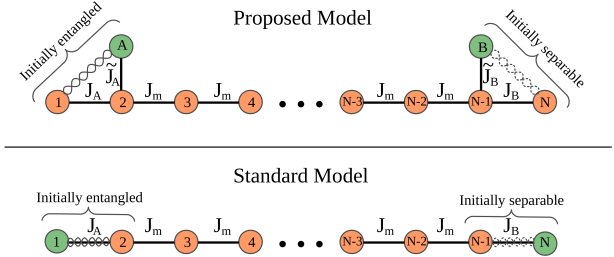}
\caption{\label{fig1}%
Upper panel: Initially qubits $A$ and $1$ with Alice are prepared by her in the maximally entangled state $|\Psi^+\rangle=(|01\rangle + |10\rangle)/\sqrt{2}$
and all the other qubits are in the state $|0\rangle$. Our goal is to find the optimal constants $J_{\!_A}$, $\tilde{J}_{\!_A}$, $J_m$, $J_{\!_B}$, $\tilde{J}_{\!_B}$, and 
time $t$ leading to the best pairwise entanglement transmission, i.e., we want the set of coupling constants and time $t$ for which Bob's two qubits $N$ and $B$ 
become most entangled.
Lower panel: Initially qubits $1$ and $2$ with Alice are prepared by her in the maximally entangled state $|\Psi^+\rangle=(|01\rangle + |10\rangle)/\sqrt{2}$
and all the other qubits are in the state $|0\rangle$. Our goal is to find the optimal constants $J_{\!_A}$, $J_m$, $J_{\!_B}$, and 
time $t$ leading to the best pairwise entanglement transmission, i.e., we want the set of coupling constants and time $t$ for which Bob's two qubits $N-1$ 
and $N$ become most entangled.
}
\end{center} \end{figure}

In almost all these 
works the main focus was the study of a strictly one dimensional graph (spin chain-like systems), which we call the standard model (see the lower panel
of Fig. \ref{fig1}). 
In the notation of the lower panel of Fig. \ref{fig1}, task (a) is related to transferring the state
describing qubit $1$ to qubit $N$ while task (b) consists in preparing qubits $1$ and $2$ in a highly entangled state and wait long
enough to obtain qubits $1$ and $N$ in a highly entangled state. Task (c), on the other hand, aims at transferring, for instance, a quantum state
describing initially qubits $1$ and $2$ to qubits $N-1$ and $N$.
See also Refs. \cite{ple04,sem05,har06,nic16} for chains built with continuous variable systems, i.e., systems described by pairs of 
canonically conjugated variables such as position and momentum.   

In this work we are not interested in the sharing of entanglement between Alice and Bob, as described in task (b) above, or the transfer
of single qubit states, i.e., task (a). Our focus is on task (c), with the following two ingredients. 
First, we are not concerned in the
transfer of arbitrary two-qubit states. We want to study the transfer of maximally entangled two-qubit states, namely, Bell states. 
In other words, our main goal here is to investigate the transmission of the pairwise entanglement between two qubits with Alice to two qubits
with Bob. The two qubits with Alice are prepared in a maximally entangled Bell state at time $t=0$ and our goal is to find
the optimal coupling constants and time $t$ leading to the greatest pairwise entanglement between two qubits with Bob (see Fig. \ref{fig1}). 
 Second, and as depicted in the upper panel of Fig. \ref{fig1}, we work \textit{slightly} beyond a one dimensional graph (spin chain). 
This geometry is crucial to have almost perfect transmission of a Bell state without a modulated chain \cite{chr04,nik04,kay05}
or external fields acting on the spins \cite{shi05,kay11,lor13,lor15}.

Indeed, as we show in the following sections, and much to our surprise, the standard model (lower panel of Fig. \ref{fig1}) gives very poor results in 
accomplishing this task for the simple unmodulated settings of Fig. \ref{fig1}, specially for long chains. 
However, the slightly modified spin chain (strictly speaking a two dimensional graph)
shown in the upper panel of Fig. \ref{fig1} gives extraordinary results, leading to an almost perfect pairwise entanglement transmission
for spin chains of any size. Also, when studying the robustness of the proposed model, we observed that \textit{small disorder}
leads to a \textit{greater} efficiency in many situations, a counter-intuitive result. 

We should also explicitly mention the interesting work of Chen \textit{et al.} \cite{che16}, where  
a similar geometry to that shown in the upper panel of Fig. \ref{fig1} is employed to transfer a single excitation from qubit $1$ to $N$ (task (a)). 
In Ref. \cite{che16} it is shown that one must couple qubits $A$ and $B$ 
with qubits $3$ and $N-2$, respectively, instead of qubits $2$ and $N-1$ as in our model, in order
to get an efficient transfer. In Ref. \cite{che16} the authors set $J_{\!_A}=J_m=J_{\!_B}=J$ and $\tilde{J}_{\!_A}=\tilde{J}_{\!_B}=w$ and search for the optimal $w/J$ leading to the
best single excitation transfer. 

Before we get to Sec. \ref{results}, where a systematic and extensive comparative study is made between the pairwise entanglement transmission 
efficiencies of the standard and proposed models, we give in Sec. \ref{tools} the mathematical formulation of the models studied here as well as other 
quantities and concepts needed to compute the efficiency of pairwise entanglement transmission. In Sec. \ref{robustness} we test the proposed model
against imperfections in its construction by studying how its efficiency is affected by static disorder. We show that for up to a moderately disordered
system we obtain very good efficiencies and still outperform the standard model. For small disorder we can even get better results than
those of the corresponding clean system. For a comprehensive analysis of the influence of noise and disorder in the functioning of the 
standard model we recommend Refs. \cite{nik04,chi05,fit05,bur05,pet10,zwi11,bru12,nik13,kur14,ash15,pav16,ron16,lyr17,lyr17a,lyr17b}. 
In Sec. \ref{single} we analyze, for completeness, 
how efficient the proposed and the standard models are in the transmission of a single excitation.
In this case, the standard model is the best choice. 
Finally, in Sec. \ref{conclusion} we give our final thoughts on the subject of 
this manuscript and also propose further lines of research we believe might enhance our 
understanding of single state and pairwise entanglement transmissions
along a spin chain under more realistic settings.

\section{The mathematical tools}
\label{tools}

In Sec. \ref{model} we give the Hamiltonian describing the systems depicted in Fig. \ref{fig1} as well as
how to efficiently solve numerically the Schr\"odinger equation that dictates their dynamics.  
In Sec. \ref{eof} we present the measure we employ to quantify the pairwise 
entanglement between two qubits and how we quantify the similarity between two quantum states.

\subsection{The model and its time evolution}
\label{model}

The Hamiltonian describing the proposed model is the isotropic XY model (XX model) with two extra
qubits $A$ and $B$ coupled with qubits $2$ and $N-1$, respectively. We have a total of $N+2$ qubits and
the Hamiltonian reads
\begin{equation}
H = H_A + H_M + H_B,
\label{ham0}
\end{equation}
where
\begin{eqnarray}
H_A  \hspace{-.2cm}&=&\hspace{-.2cm}  J_{\!_A} (\sigma_1^x\sigma_2^x+\sigma_1^y\sigma_2^y) + \tilde{J}_{\!_A} (\sigma_A^x\sigma_2^x+\sigma_A^y\sigma_2^y), \nonumber \\
H_M  \hspace{-.2cm}&=&\hspace{-.2cm}  \sum_{j=2}^{N-2}J_m(\sigma_j^x\sigma_{j+1}^x+\sigma_j^y\sigma_{j+1}^y), \nonumber \\
H_B  \hspace{-.2cm}&=&\hspace{-.2cm}  J_{\!_B} (\sigma_{N-1}^x\sigma_N^x\!+\!\sigma_{N-1}^y\sigma_N^y) \!+\! 
\tilde{J}_{\!_B} (\sigma_{N-1}^x\sigma_B^x\!+\!\sigma_{N-1}^y\sigma_B^y). \nonumber \\
\label{ham}
\end{eqnarray}
Here $\sigma_i^\alpha\sigma_j^\alpha = \sigma_i^\alpha \otimes \sigma_j^\alpha$, with the superscript denoting a particular Pauli matrix 
and the subscript labeling the qubit acted by it. 
We also adopt the following convention usually employed in the quantum information community: 
$\sigma^z|0\rangle=|0\rangle,\sigma^z|1\rangle=-|1\rangle,
\sigma^x|0\rangle=|1\rangle,\sigma^x|1\rangle=|0\rangle,\sigma^y|0\rangle=i|1\rangle,\sigma^y|1\rangle=-i|0\rangle$, where $i$ is the imaginary unity
and $|0\rangle$ and $|1\rangle$ are the eigenvectors of $\sigma^z$. For those used to the up and down notation of the condensed matter physics community, the relation
between the latter and the present notation is $|\uparrow\rangle=|0\rangle$ and $|\downarrow\rangle = |1\rangle$. Note that if we set $\tilde{J}_{\!_A}=\tilde{J}_{\!_B}=0$ we 
get the Hamiltonian describing the standard model.

An important property of Hamiltonian (\ref{ham}) is that it commutes with the operator $Z=\sigma_A^z+ \sum_{j=1}^{N}\sigma_j^z+\sigma_B^z$, 
namely, $[H,Z]=0$. This means that the total number of excitations (spins up and down) are conserved during the time evolution of the system,
allowing us to solve the Schr\"odinger equation by restricting 
ourselves to the subspace spanned by all states with the same number of excitations of the initial state. 
This is the reason why we can numerically and efficiently study the pairwise entanglement transmission for chains of the order of $1000$ qubits, 
specially if we deal with states with only one excitation such as the Bell state $|\Psi^+\rangle=(|01\rangle+|10\rangle)/\sqrt{2}$.

From now on we restrict ourselves to the sector of one excitation and to the labeling of the qubits 
as given in Fig. \ref{fig1}. In this scenario, an arbitrary system composed of $N+2$ qubits 
can be described by the superposition of $N+2$ states of one excitation, 
\begin{equation}
|\Psi(t)\rangle = \sum_j c_{\!_j}(t)|1_j\rangle,
\label{psit}
\end{equation}
where $j=A,1,2, \ldots, N-1,N,B$, and
%
\begin{equation}
|1_j\rangle = \sigma_j^x|000\cdots\hspace{-.25cm} \underbrace{0}_\text{j-th qubit}\hspace{-.25cm} \cdots 000\rangle 
 = |000\cdots\hspace{-.25cm} \underbrace{1}_\text{j-th qubit}\hspace{-.25cm} \cdots 000\rangle. 
\end{equation}

If at time $t=0$ qubits $1$ and $A$ are given by the Bell state $|\Psi^+\rangle$, i.e., 
$|\Psi(0)\rangle=(|010\cdots 0\rangle+|10\cdots 0\rangle)/\sqrt{2}$, the initial conditions in the notation 
of Eq.~(\ref{psit}) are
\begin{eqnarray}
c_{\!_A}(0)&=&c_{\!_1}(0)=1/\sqrt{2}, 
\label{initial1}\\
c_{\!_j}(0)&=&0, \hspace{.1cm} \mbox{for} \hspace{.1cm} j \neq A,1.
\label{initial2}
\end{eqnarray}
For the standard model we have $j=1,\ldots,N$, $c_{\!_1}(0)=c_{\!_2}(0)=1/\sqrt{2}$, and
$c_{\!_j}(0)=0, \hspace{.1cm} \mbox{for} \hspace{.1cm} j \neq 1,2$.

Inserting Eq.~(\ref{psit}) into the Schr\"odinger equation 
$$
i\hbar \frac{d|\Psi(t)\rangle}{dt} = H |\Psi(t)\rangle
$$ 
we get after taking the scalar product with the bra $\langle 1_k|$,
\begin{equation}
i\hbar \frac{dc_{\!_k}(t)}{dt} = \sum_j c_{\!_j}(t) \langle 1_k|H|1_j\rangle.
\label{ct}
\end{equation}

A long but direct calculation leads to
\begin{eqnarray}
\langle 1_k|H|1_j\rangle & = & 2 J_{\!_A}(\delta_{1k}\delta_{j2}+\delta_{2k}\delta_{j1}) \nonumber \\
&& + 2\tilde{J}_{\!_A}(\delta_{Ak}\delta_{j2}+\delta_{2k}\delta_{jA}) \nonumber \\
&& + 2 J_{\!_B}(\delta_{N-1,k}\delta_{jN}+\delta_{Nk}\delta_{j,N-1}) \nonumber \\
&& + 2 \tilde{J}_{\!_B}(\delta_{N-1,k}\delta_{jB}+\delta_{Bk}\delta_{j,N-1}) \nonumber \\
&& + 2J_m \sum_{l=2}^{N-2}(\delta_{lk}\delta_{j,l+1}+\delta_{l+1,k}\delta_{jl}),
\label{braket}
\end{eqnarray}
where the Kronecker delta is defined such that $\delta_{jk}=1$ if $j=k$ and $\delta_{jk}=0$ if $j\neq k$. 

Now, inserting Eq.~(\ref{braket}) into (\ref{ct}) gives
\begin{eqnarray}
\frac{dc_{\!_k}(t)}{dt} & = & -i\frac{2}{\hbar}\tilde{J}_{\!_A}\delta_{2k}c_{\!_A}(t) - i\frac{2}{\hbar}J_{\!_A}\delta_{2k}c_{\!_1}(t) \nonumber \\
& & -i\frac{2}{\hbar}(J_{\!_A}\delta_{1k}+\tilde{J}_{\!_A}\delta_{Ak}+J_m\delta_{3k})c_{\!_2}(t) \nonumber \\
& & -i\frac{2}{\hbar}\sum_{j=3}^{N-2}(J_m\delta_{j-1,k}+J_m\delta_{j+1,k})c_{\!_j}(t) \nonumber \\
& & -i\frac{2}{\hbar}(J_m\delta_{N-2,k}+\tilde{J}_{\!_B}\delta_{Bk}+J_{\!_B}\delta_{Nk})c_{\!_{N-1}}(t) \nonumber \\
& & -i\frac{2}{\hbar}J_{\!_B}\delta_{N-1,k}c_{\!_N}(t) - i\frac{2}{\hbar}\tilde{J}_{\!_B}\delta_{N-1,k}c_{\!_B}(t).
\label{difeq}
\end{eqnarray}

Equation (\ref{difeq}) is a system of $N+2$ first order linear differential equations with constant 
coefficients, where
$k=A,1,\ldots,N,B$. 

Defining the column vector
\begin{equation}
\mathbf{c}(t) =
\left(
\begin{array}{c}
c_{\!_A}(t) \\
c_{\!_1}(t) \\
c_{\!_2}(t) \\
\vdots \\
c_{\!_{N-1}}(t) \\
c_{\!_N}(t) \\
c_{\!_B}(t)
\end{array}
\right)
\end{equation}
and the matrix 
\begin{equation}
\mathbf{M} \hspace{-.1cm}=\hspace{-.1cm} -\frac{i2}{\hbar}\hspace{-.1cm}
\left(\hspace{-.2cm}
\begin{array}{cccccccccccc}
0 \hspace{-.2cm} & \hspace{-.2cm} 0 \hspace{-.2cm} & \hspace{-.2cm} \tilde{J}_{\!_A} \hspace{-.2cm} & \hspace{-.2cm} 0 \hspace{-.2cm} & \hspace{-.2cm} 0 \hspace{-.2cm} & \hspace{-.2cm} 0 \hspace{-.2cm} & \hspace{-.2cm} 0 \hspace{-.2cm} & \hspace{-.2cm}\cdots \hspace{-.2cm} & \hspace{-.2cm} 0 \hspace{-.2cm} & \hspace{-.2cm} 0 \hspace{-.2cm} & \hspace{-.2cm} 0 \hspace{-.2cm} & \hspace{-.2cm} 0\\
0 \hspace{-.2cm} & \hspace{-.2cm} 0 \hspace{-.2cm} & \hspace{-.2cm} J_{\!_A} \hspace{-.2cm} & \hspace{-.2cm} 0 \hspace{-.2cm} & \hspace{-.2cm} 0 \hspace{-.2cm} & \hspace{-.2cm} 0 \hspace{-.2cm} & \hspace{-.2cm} 0 \hspace{-.2cm} & \hspace{-.2cm} \cdots \hspace{-.2cm} & \hspace{-.2cm} 0 \hspace{-.2cm} & \hspace{-.2cm} 0 \hspace{-.2cm} & \hspace{-.2cm} 0 \hspace{-.2cm} & \hspace{-.2cm} 0\\
\tilde{J}_{\!_A} \hspace{-.2cm} & \hspace{-.2cm} J_{\!_A} \hspace{-.2cm} & \hspace{-.2cm} 0 \hspace{-.2cm} & \hspace{-.2cm} J_m \hspace{-.2cm} & \hspace{-.2cm} 0 \hspace{-.2cm} & \hspace{-.2cm} 0 \hspace{-.2cm} & \hspace{-.2cm} 0 \hspace{-.2cm} & \hspace{-.2cm} \cdots \hspace{-.2cm} & \hspace{-.2cm} 0 \hspace{-.2cm} & \hspace{-.2cm} 0 \hspace{-.2cm} & \hspace{-.2cm} 0 \hspace{-.2cm} & \hspace{-.2cm} 0\\
0 \hspace{-.2cm} & \hspace{-.2cm} 0 \hspace{-.2cm} & \hspace{-.2cm} J_m \hspace{-.2cm} & \hspace{-.2cm} 0 \hspace{-.2cm} & \hspace{-.2cm} J_m \hspace{-.2cm} & \hspace{-.2cm} 0 \hspace{-.2cm} & \hspace{-.2cm} 0 \hspace{-.2cm} & \hspace{-.2cm} \cdots  \hspace{-.2cm} & \hspace{-.2cm} 0 \hspace{-.2cm} & \hspace{-.2cm} 0 \hspace{-.2cm} & \hspace{-.2cm} 0 \hspace{-.2cm} & \hspace{-.2cm} 0\\
0 \hspace{-.2cm} & \hspace{-.2cm} 0 \hspace{-.2cm} & \hspace{-.2cm} 0 \hspace{-.2cm} & \hspace{-.2cm} J_m \hspace{-.2cm} & \hspace{-.2cm} 0 \hspace{-.2cm} & \hspace{-.2cm} J_m \hspace{-.2cm} & \hspace{-.2cm} 0 \hspace{-.2cm} & \hspace{-.2cm}\cdots  \hspace{-.2cm} & \hspace{-.2cm} 0 \hspace{-.2cm} & \hspace{-.2cm} 0 \hspace{-.2cm} & \hspace{-.2cm} 0 \hspace{-.2cm} & \hspace{-.2cm} 0\\
0 \hspace{-.2cm} & \hspace{-.2cm} 0 \hspace{-.2cm} & \hspace{-.2cm} 0 \hspace{-.2cm} & \hspace{-.2cm} 0 \hspace{-.2cm} & \hspace{-.2cm} J_m \hspace{-.2cm} & \hspace{-.2cm} 0 \hspace{-.2cm} & \hspace{-.2cm} J_m \hspace{-.2cm} & \hspace{-.2cm}\cdots  \hspace{-.2cm} & \hspace{-.2cm} 0 \hspace{-.2cm} & \hspace{-.2cm} 0 \hspace{-.2cm} & \hspace{-.2cm} 0 \hspace{-.2cm} & \hspace{-.2cm} 0\\
\vdots \hspace{-.2cm} & \hspace{-.2cm} \vdots \hspace{-.2cm} & \hspace{-.2cm} \vdots \hspace{-.2cm} & \hspace{-.2cm} \vdots \hspace{-.2cm} & \hspace{-.2cm} \vdots \hspace{-.2cm} & \hspace{-.2cm} \vdots \hspace{-.2cm} & \hspace{-.2cm} \vdots \hspace{-.2cm} & \hspace{-.2cm}\ddots  \hspace{-.2cm} & \hspace{-.2cm} \vdots \hspace{-.2cm} & \hspace{-.2cm} \vdots \hspace{-.2cm} & \hspace{-.2cm} \vdots \hspace{-.2cm} & \hspace{-.2cm} \vdots\\
0 \hspace{-.2cm} & \hspace{-.2cm} 0 \hspace{-.2cm} & \hspace{-.2cm} 0 \hspace{-.2cm} & \hspace{-.2cm} \cdots \hspace{-.2cm} & \hspace{-.2cm} 0 \hspace{-.2cm} & \hspace{-.2cm} 0 \hspace{-.2cm} & \hspace{-.2cm} J_m \hspace{-.2cm} & \hspace{-.2cm} 0 \hspace{-.2cm} & \hspace{-.2cm} J_m \hspace{-.2cm} & \hspace{-.2cm} 0 \hspace{-.2cm} & \hspace{-.2cm} 0 \hspace{-.2cm} & \hspace{-.2cm} 0 \\
0 \hspace{-.2cm} & \hspace{-.2cm} 0 \hspace{-.2cm} & \hspace{-.2cm} 0 \hspace{-.2cm} & \hspace{-.2cm} \cdots \hspace{-.2cm} & \hspace{-.2cm} 0 \hspace{-.2cm} & \hspace{-.2cm} 0 \hspace{-.2cm} & \hspace{-.2cm} 0 \hspace{-.2cm} & \hspace{-.2cm} J_m \hspace{-.2cm} & \hspace{-.2cm} 0 \hspace{-.2cm} & \hspace{-.2cm} J_m \hspace{-.2cm} & \hspace{-.2cm} 0 \hspace{-.2cm} & \hspace{-.2cm} 0 \\
0 \hspace{-.2cm} & \hspace{-.2cm} 0 \hspace{-.2cm} & \hspace{-.2cm} 0 \hspace{-.2cm} & \hspace{-.2cm} \cdots \hspace{-.2cm} & \hspace{-.2cm} 0 \hspace{-.2cm} & \hspace{-.2cm} 0 \hspace{-.2cm} & \hspace{-.2cm} 0 \hspace{-.2cm} & \hspace{-.2cm} 0 \hspace{-.2cm} & \hspace{-.2cm} J_m \hspace{-.2cm} & \hspace{-.2cm} 0 \hspace{-.2cm} & \hspace{-.2cm} J_{\!_B} \hspace{-.2cm} & \hspace{-.2cm} \tilde{J}_{\!_B} \\
0 \hspace{-.2cm} & \hspace{-.2cm} 0 \hspace{-.2cm} & \hspace{-.2cm} 0 \hspace{-.2cm} & \hspace{-.2cm} \cdots \hspace{-.2cm} & \hspace{-.2cm} 0 \hspace{-.2cm} & \hspace{-.2cm} 0 \hspace{-.2cm} & \hspace{-.2cm} 0 \hspace{-.2cm} & \hspace{-.2cm} 0 \hspace{-.2cm} & \hspace{-.2cm} 0 \hspace{-.2cm} & \hspace{-.2cm} J_{\!_B} \hspace{-.2cm} & \hspace{-.2cm} 0 \hspace{-.2cm} & \hspace{-.2cm} 0 \\
0 \hspace{-.2cm} & \hspace{-.2cm} 0 \hspace{-.2cm} & \hspace{-.2cm} 0 \hspace{-.2cm} & \hspace{-.2cm} \cdots \hspace{-.2cm} & \hspace{-.2cm} 0 \hspace{-.2cm} & \hspace{-.2cm} 0 \hspace{-.2cm} & \hspace{-.2cm} 0 \hspace{-.2cm} & \hspace{-.2cm} 0 \hspace{-.2cm} & \hspace{-.2cm} 0 \hspace{-.2cm} & \hspace{-.2cm} \tilde{J}_{\!_B} \hspace{-.2cm} & \hspace{-.2cm} 0 \hspace{-.2cm} & \hspace{-.2cm} 0
\end{array}
\hspace{-.2cm}\right)
\label{matriz}
\end{equation}
we can recast Eq.~(\ref{difeq}) as
\begin{equation}
\frac{d\mathbf{c}(t)}{dt} = \mathbf{M} \,\, \mathbf{c}(t).
\label{mateq}
\end{equation}
The formal solution to Eq.~(\ref{mateq}) is
\begin{equation}
\mathbf{c}(t) = \exp(\mathbf{M}\,\, t )\mathbf{c}(0),
\end{equation}
where ``$\exp$'' stands for the matrix exponential and  
$\mathbf{c}(0)$ for the column vector listing the initial conditions as 
given in Eqs.~(\ref{initial1}) and (\ref{initial2}).

It is important to note that $\mathbf{M}$ is an $(N+2) \times (N+2)$ dimensional matrix and
by using efficient numerical packages already available to compute the matrix exponential we
can solve for chains of about $1000$ qubits using off the shelf desktop computers.

\subsection{Entanglement and Fidelity}
\label{eof}

\subsubsection{Entanglement of formation}

The key mathematical object needed in the analysis of the pairwise entanglement 
between qubits $N$ and $B$ with Bob is their density matrix $\rho_{\!_{NB}}(t)$, which is obtained
by tracing out the other $N$ qubits from the density matrix $\rho(t)$ describing the whole system,
\begin{equation}
\rho_{\!_{NB}}(t) = \mbox{Tr}_{\overline{NB}} [\rho(t)].
\end{equation}
The line over NB is a reminder that we are tracing out all but qubits $N$ and $B$ from $\rho(t)$.
The density matrix of the whole system is $\rho(t)=|\Psi(t)\rangle\langle\Psi(t)|$, where 
$|\Psi(t)\rangle$ is given by Eq.~(\ref{psit}). 

Remembering that we are working within the one excitation
subspace and using the normalization condition $\sum_j|c_{\!_j}(t)|^2=1$, a simple calculation leads to
\begin{eqnarray}
\hspace{-.15cm}\rho_{\!_{NB}}(t) \hspace{-.15cm}&=&\hspace{-.15cm} (1-|c_{\!_N}(t)|^2-|c_{\!_B}(t)|^2)|00\rangle\langle 00| \nonumber \\
\hspace{-.15cm}& + &\hspace{-.15cm} |c_{\!_B}(t)|^2|01\rangle\langle 01| + |c_{\!_N}(t)|^2|10\rangle\langle 10| \nonumber \\
\hspace{-.15cm}& + &\hspace{-.15cm} c_{\!_B}(t)c_{\!_N}^*(t)|01\rangle\langle 10| + c_{\!_N}(t)c_{\!_B}^*(t)|10\rangle\langle 01| \nonumber \\
\hspace{-.15cm}&=&\hspace{-.25cm} 
\left(\hspace{-.12cm}
\begin{array}{cccc}
1-|c_{\!_N}(t)|^2-|c_{\!_B}(t)|^2 \hspace{-.125cm}&\hspace{-.125cm} 0 \hspace{-.125cm}&\hspace{-.125cm} 0 & 0 \\
0 \hspace{-.125cm}&\hspace{-.125cm} |c_{\!_B}(t)|^2 \hspace{-.125cm}&\hspace{-.125cm} c_{\!_B}(t)c_{\!_N}^*(t) & 0 \\
0 \hspace{-.125cm}&\hspace{-.125cm} c_{\!_N}(t)c_{\!_B}^*(t) \hspace{-.125cm}&\hspace{-.125cm} |c_{\!_N}(t)|^2 & 0 \\
0 \hspace{-.125cm}&\hspace{-.125cm} 0 \hspace{-.125cm}&\hspace{-.125cm} 0 & 0
\end{array}
\hspace{-.12cm}
\right), \nonumber \\
\label{rhoNB}
\end{eqnarray}
where $*$ means complex conjugation. Equation (\ref{rhoNB}) tells us that we only need to know the time
dependence of the two coefficients $c_{\!_N}(t)$ and $c_{\!_B}(t)$ to completely describe the two qubits 
$N$ and $B$ with Bob. For the standard model the density matrix describing qubits $N-1$ and $N$ is given
by Eq.~(\ref{rhoNB}) with $c_{\!_N}(t)$ and $c_{\!_B}(t)$ changed to $c_{\!_{N-1}}(t)$ and $c_{\!_N}(t)$.

The density matrix describing qubit $B$, which will be used in the analysis of
the single excitation transmission in Sec. \ref{single}, is obtained from Eq.~(\ref{rhoNB}) by tracing out
qubit $N$ and can be written as
\begin{eqnarray}
\hspace{-.15cm}\rho_{\!_{B}}(t) 
\hspace{-.15cm}&=&\hspace{-.25cm} 
\left(\hspace{-.12cm}
\begin{array}{cc}
1-|c_{\!_B}(t)|^2 \hspace{-.125cm}&\hspace{-.125cm} 0  \\
0 \hspace{-.125cm}&\hspace{-.125cm} |c_{\!_B}(t)|^2
\end{array}
\hspace{-.12cm}
\right). 
\label{rhoB}
\end{eqnarray}
In the standard model instead of $\rho_{\!_{B}}(t)$ we need $\rho_{\!_{N}}(t)$, which is given
by Eq.~(\ref{rhoB}) with $c_{\!_B}(t)$ changed to $c_{\!_N}(t)$.

The pairwise entanglement between two qubits can be computed by an
analytical expression called Entanglement of Formation (EoF) \cite{ben96,woo98}. If 
$\rho_{\!_{NB}}$ is the density matrix describing our two qubits, the EoF is
obtained by minimizing the average of the entanglement of the pure state decomposition of
$\rho_{\!_{NB}}$over all possible decompositions,
\begin{equation}
 \mbox{EoF}(\rho_{\!_{NB}}) = \min_{\{p_k,|\phi_k\rangle\}}\sum_{k}p_{k}E(|\phi_{k}\rangle), 
\end{equation}
where $\sum_{k}p_{k} = 1$, $0 < p_{k} \leq 1$, and $\rho_{\!_{NB}} =
\sum_{k}p_{k}\left| \phi_{k}\right>\left< \phi_{k}\right|$ is a possible decomposition 
of $\rho_{\!_{NB}}$.
$E(|\phi_{k}\rangle)$ is the entanglement of the pure state
$|\phi_{k}\rangle$ quantified by the von Neumann entropy of the single qubit
state obtained by tracing out the other qubit from $|\phi_{k}\rangle$ \cite{ben96b}. 

For two qubits Wootters
\cite{woo98} proved that the EoF is a monotonically increasing
function of the concurrence $C$ and that 
\begin{equation}
\mbox{EoF}(\rho_{\!_{NB}}) = -f(C)\log_2f(C) - [1-f(C)]\log_2[1-f(C)], 
\end{equation}
with $f(C)=(1 + \sqrt{1 - C^2})/2$. A maximally entangled state has
EoF $= 1.0$ and a completely unentangled system has EoF $= 0$. 

The concurrence is given by \cite{woo98}
\begin{equation}
C = \mbox{max} \{ 0, \lambda_{1} - \lambda_{2} - \lambda_{3} -
\lambda_{4} \},
\label{concurrence}
\end{equation}
where $\lambda_{1},  \lambda_{2}, \lambda_{3}$, and $\lambda_{4}$
are, in decreasing order, the square roots of the eigenvalues of
the matrix $R =\rho_{\!_{NB}}\left(\sigma^{y}_N \otimes \sigma^{y}_B \right)
\rho_{\!_{NB}}^{*} \left(\sigma^{y}_N \otimes \sigma^{y}_B \right)$.
Here $\rho_{\!_{NB}}^{*}$ stands for the complex conjugation of
$\rho_{\!_{NB}}$ in the standard basis $\left\{ \left| 00 \right>, \left| 01
\right>, \left| 10 \right>, \left| 11 \right> \right\}$. 

For a density matrix in the form of Eq.~(\ref{rhoNB}), a direct calculation gives
\begin{equation}
C(t) = 2|c_{\!_N}(t)c_{\!_B}(t)|.
\label{concurrence2}
\end{equation}


\subsubsection{Fidelity}

In order to quantify in a simple way the ``closeness'' between two quantum states, in particular when the 
benchmark or input state is a pure one, we employ the fidelity
\begin{equation}
F = \langle \psi_{in}| \rho_{out} | \psi_{in}\rangle.
\label{fid}
\end{equation}
If the input state is given by the Bell state $|\Psi^+\rangle=(|01\rangle+|10\rangle)/\sqrt{2}$
and the output state is the two qubit state $\rho_{\!_{NB}}$, as given by Eq.~(\ref{rhoNB}), 
we have
\begin{equation}
F(t) = |c_{\!_N}(t)+c_{\!_B}(t)|^2/2. 
\end{equation}

At the single excitation level, where the input state is $|1\rangle$ and the output state is the 
single qubit given by Eq.~(\ref{rhoB}), we get $F=|c_{\!_B}(t)|^2$. 

The fidelity between two
states differing only by an overall global phase is one while it is zero for two orthogonal states.
Both the EoF and the fidelity ranges from zero to one. The greater the value of  
the EoF the more a two-qubit state
is entangled and the greater the value of the fidelity the more similar are two quantum states.

\section{Pairwise entanglement transmission}
\label{results}

\begin{figure*}[!ht]
\begin{center}
\includegraphics[width=10cm,height=5cm]{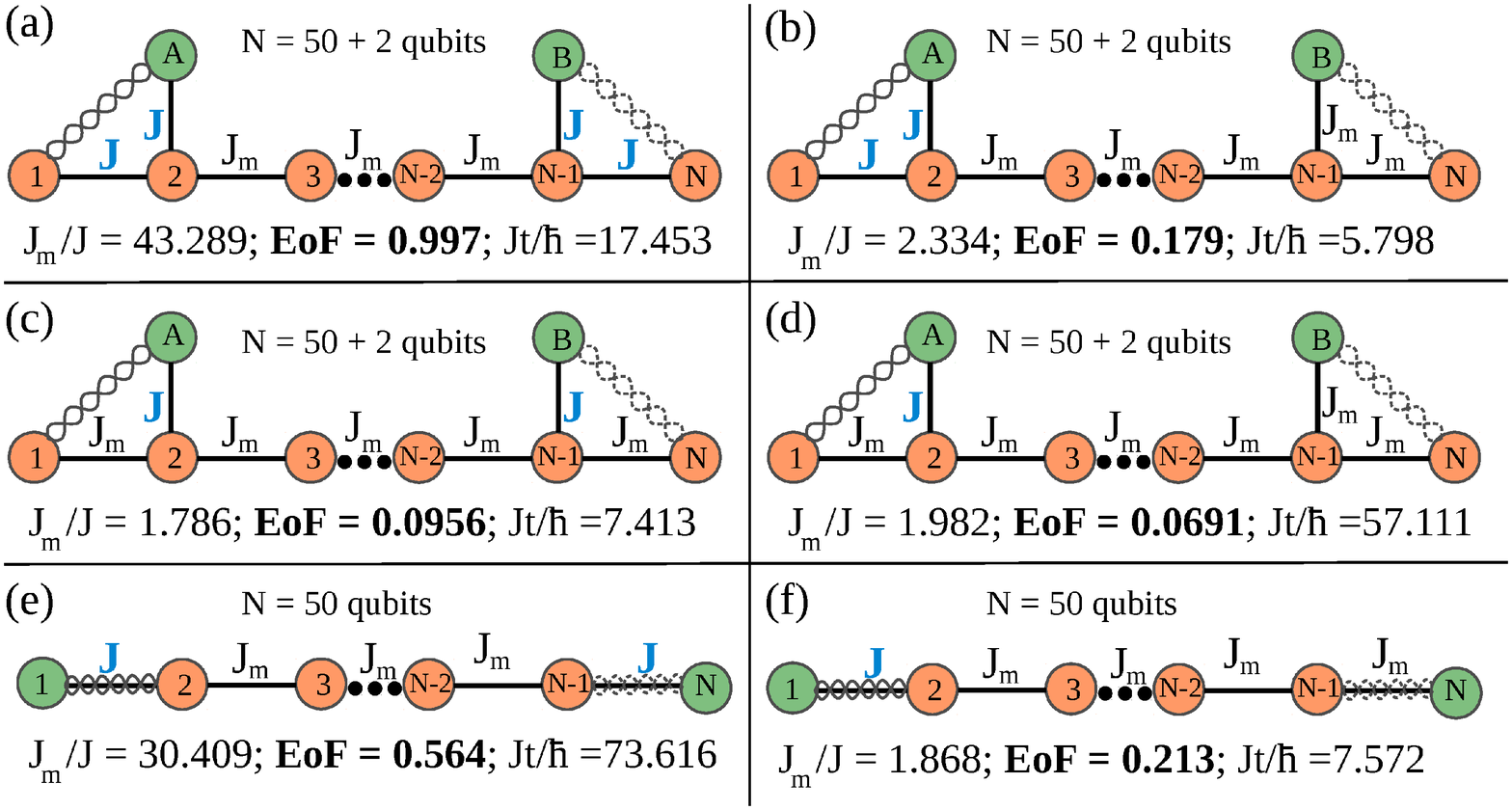}
\caption{\label{fig2}%
Panels (a) to (d) show four possible relationships among the coupling constants for the proposed model and 
panels (e) and (f) two configurations for the standard model. For all cases we have changed 
$J_m/J$ from $-50$ to $50$ and the system always started with the two qubits with Alice given by
the maximally entangled state $|\Psi^+\rangle=(|01\rangle + |10\rangle)/\sqrt{2}$ and the remaining qubits
in the state $|0\rangle$. For each value of $J_m/J$ we have evolved the system up to the dimensionless 
time $Jt/\hbar = 25\pi\approx 78.54$, computing along the way the entanglement of formation (EoF)
between qubits $N$ and $B$ (or $N-1$ and $N$ for panels (e) and (f)) as a function of $Jt/\hbar$. 
The data shown in panels (a)-(f) are those corresponding to the greatest EoF between the qubits with Bob
within the range of values for $J_m/J$ and $Jt/\hbar$ as specified above.
Note that the configuration depicted in panel (a) stands out among the other ones, leading to 
almost perfect entanglement transmission (EoF $= 0.997$). 
}
\end{center}
\end{figure*}

The model we are dealing with, Eqs.~(\ref{ham0}) and (\ref{ham}), 
has five independent coupling constants, namely,
$J_{\!_A}$, $\tilde{J}_{\!_A}$, $J_m$, $J_{\!_B}$, $\tilde{J}_{\!_B}$.
Our first task is to search for an optimal relationship among these
coupling constants that leads to the greatest efficiency in the transmission of
pairwise entanglement from qubits $A$ and $1$ to qubits $N$ and $B$ 
(or from $1$ and $2$ to $N-1$ and $N$ in the standard model). 

We tested several arrangements for systems with $N=10,20,30,40,$ and $50$ qubits.
For all simulations we obtained a similar qualitative behavior, i.e., the proposed
model works best in the transmission of pairwise entanglement whenever 
$J_{\!_A} = \tilde{J}_{\!_A} = J_{\!_B} = \tilde{J}_{\!_B}=J$ and $J_m\neq J$.
We also observed that the greater $N$ the more the optimal configuration for the proposed model stands out
as the only configuration leading to a meaningful entanglement transmission. The standard model
optimal configuration, on the other hand, is given by  
$J_{\!_A} =  J_{\!_B} = J$ and $J_m\neq J$, leading, however, to a less efficient transmission.
In Fig. \ref{fig2} we show the quantitative results for $N=50$ qubits. 

It is worth noting that by applying the optimal arrangement among 
the coupling constants in the proposed model, Hamiltonian (\ref{ham}) 
becomes invariant if we interchange qubits 
$A$ with $1$ or qubits $N$ with $B$. 
The latter symmetry and the fact that the initial state $|\Psi^+\rangle\otimes |0\rangle^{\otimes N}$ 
is also symmetric with respect to the interchange of $N$ with $B$ imply that $c_{\!_N}(t)=c_{\!_B}(t)$. 
This last result means that the concurrence and the fidelity, Eqs.~(\ref{concurrence2}) and (\ref{fid}), are identical,
\begin{equation}
C(t) = F(t) = 2|c_{\!_B}(t)|^2. 
\end{equation}

Once we have determined the optimal configuration among the coupling constants leading to the greatest pairwise
entanglement transmission for the proposed and standard  models,
panels (a) and (e) of Fig. \ref{fig2}, respectively, 
we investigated for how long a chain we can
use those arrangements in such a way to implement an efficient pairwise entanglement transmission.

To that aim we have simulated the time evolution of chains composed of $N=100$ to $N=1000$ qubits, following 
a similar strategy to that explained in the caption of Fig. \ref{fig2}. Here, however, 
we worked with a longer time and a greater range of values for the coupling constants to cope with
the longer size of the chains.

The results we obtained pointed in the direction of a very inefficient transmission if we use 
the standard model. We have seen that the efficiency in this case decreases
with the size of the chain, giving very poor results for long chains. On the other hand,
and surprisingly, the proposed model showed no decrease in its transmission efficiency of 
pairwise entanglement. In this last case, we could always find a time and coupling constants in which
the entanglement of formation between Bob's two qubits became greater than $0.99$.

\begin{figure}[!ht]
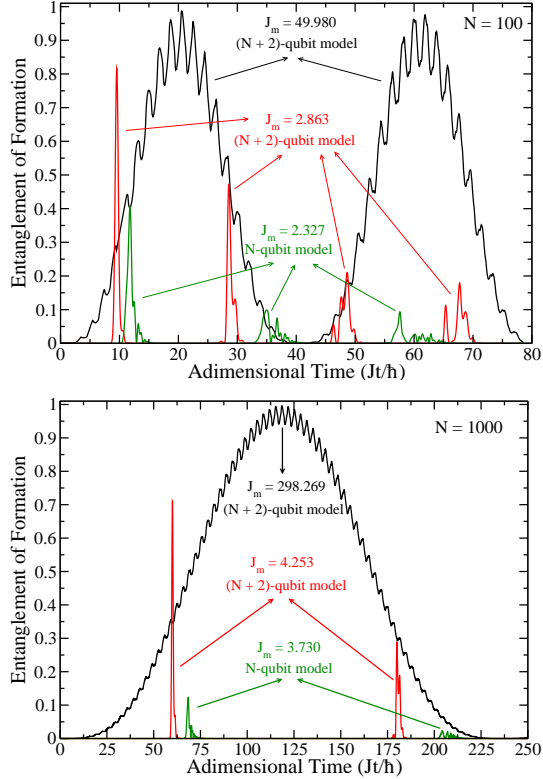
 \begin{center}
\includegraphics[width=7cm]{N100Qubits.eps}\vspace{.15cm}
\includegraphics[width=7cm]{N1000Qubits.eps}
\caption{\label{fig3e4}%
Upper panel: Entanglement transmission from qubits $A$ and $1$ ($1$ and $2$) with Alice
to Bob's two qubits
$N$ and $B$ ($N-1$ and $N$) for the proposed (standard) model when $N=100$ qubits.
The two curves with the greatest peaks correspond to the proposed model ($(N+2)$-qubit model) while
the curve with the lowest peaks is related to the standard model ($N$-qubit model). 
Alice's qubits are initially prepared 
in the maximally entangled state $|\Psi^+\rangle=(|01\rangle + |10\rangle)/\sqrt{2}$ while
all the other qubits start at the state $|0\rangle$.  
For simulation purposes, we set $J_{\!_A}=\tilde{J}_{\!_A}=J_{\!_B}=\tilde{J}_{\!_B}=J=1$ for
the proposed model and $J_{\!_A}=J_{\!_B}=J=1$ for the standard model. In both models we evolved the system from
$Jt/\hbar=0$ to $25\pi$ for every value of $J_m$ ranging from $0$ to $50$ in increments of $0.01$,
saving the value of $J_m$ giving the best entanglement transmission within the time span
given above. Subsequently we moved about this value
in increments of $0.001$ in order to get the optimal
$J_m$ shown in the figure. For the proposed model we also give the optimal $J_m$ restricting
its range from $0$ to $5$ (red curve showing the first peak above).
Lower panel: Same as upper panel  but now we have $N=1000$ qubits and we evolved the system from
$Jt/\hbar=0$ to $80\pi$ for every value of $J_m$ ranging initially from $0$ to $300$ in
increments of $0.01$.
Here and in the following figures all data are dimensionless.
}
\end{center}
\end{figure}

In the upper and lower panels of Fig. \ref{fig3e4} 
we show the quantitative results for chains of $N=100$ and
$N=1000$ qubits, respectively. Looking at those figures it becomes clear that the standard model
is already very inefficient for chains of the order of $100$ qubits. In this scenario the entanglement
transmitted is less than $0.5$ and for the case of $1000$ qubits we have the two qubits with Bob with an 
EoF $\approx 0.12$, a very poor result.

The proposed model, on the other hand, gives an EoF $\approx 0.99$ for
both the $100$ and $1000$ chains. Similar results (not shown here) were obtained for chains greater than $1000$ qubits,
although the simulation time for those cases were much longer and less numerically accurate than the ones reported here.
In all simulations we made, we could always find a value of $J_m$ and a time long enough such that the entanglement 
transmitted was of the order of $0.99$. Even without an analytical proof, we have enough numerical data to
conjecture that this trend will continue for any size of the chain, rendering the proposed model an almost perfect pairwise
entanglement transmitter for chains of arbitrary size.

We have also studied the scenario in which the range in which we search for the optimal $J_m$ is not too wide.
Restricting ourselves to the case in which $J_m$ is at most five times greater than $J$, we still
have excellent results working with the proposed model. For chains of the order of $100$ qubits
we get an entanglement transmitted of about $0.82$ and in the $1000$ qubit case we get an 
EoF $\approx 0.70$. The standard model transmits only an EoF $\approx 0.40$ and $0.12$ for chains
of $100$ and $1000$ qubits, respectively.

It is important at this point to note that we have also simulated the 
entanglement transmission using the other three Bell states. 
First, the state $|\Psi^-\rangle=(|01\rangle - |10\rangle)/\sqrt{2}$
is not transmitted at all in the proposed model.
This can be proved analytically by noting that the commutator 
$[H,|\Psi^-\rangle\langle \Psi^-| \otimes |0\rangle\langle 0|^{\otimes N}]$
is zero. In other words, the initial state $|\Psi^-\rangle\otimes|0\rangle^{\otimes N}$ 
is an eigenstate of the Hamiltonian (stationary state) and its time evolution is simply given 
by a global phase change that does not affect the probability distribution associated to the initial state.
Second, the states $|\Phi^{\pm}\rangle=(|00\rangle\pm |11\rangle)/\sqrt{2}$ are also better 
transmitted using the proposed model
as compared to the standard model. In this scenario, however, we could simulate up to chains of the order of $30$ qubits since
the subspace of two excitations leads to a matrix $\mathbf{M}$, Eq.~(\ref{matriz}), of dimension  
$N^2 \times N^2$ against a matrix of dimension $N \times N$ for the single excitation sector.
Finally, if we can apply local unitary gates to at least one of Alice's two qubits we can always convert any Bell
state with her to $|\Psi^+\rangle$. In this sense, we can always transmit any Bell state using the model 
proposed here by transforming a particular Bell state to the state $|\Psi^+\rangle$ and transmitting it to Bob, 
who subsequently transforms it
back to the desired Bell state by applying the appropriate unitary operation in one of his two qubits.

\section{Robustness of the proposed model}
\label{robustness}

\begin{figure}[!ht] \begin{center}
\includegraphics[width=8cm]{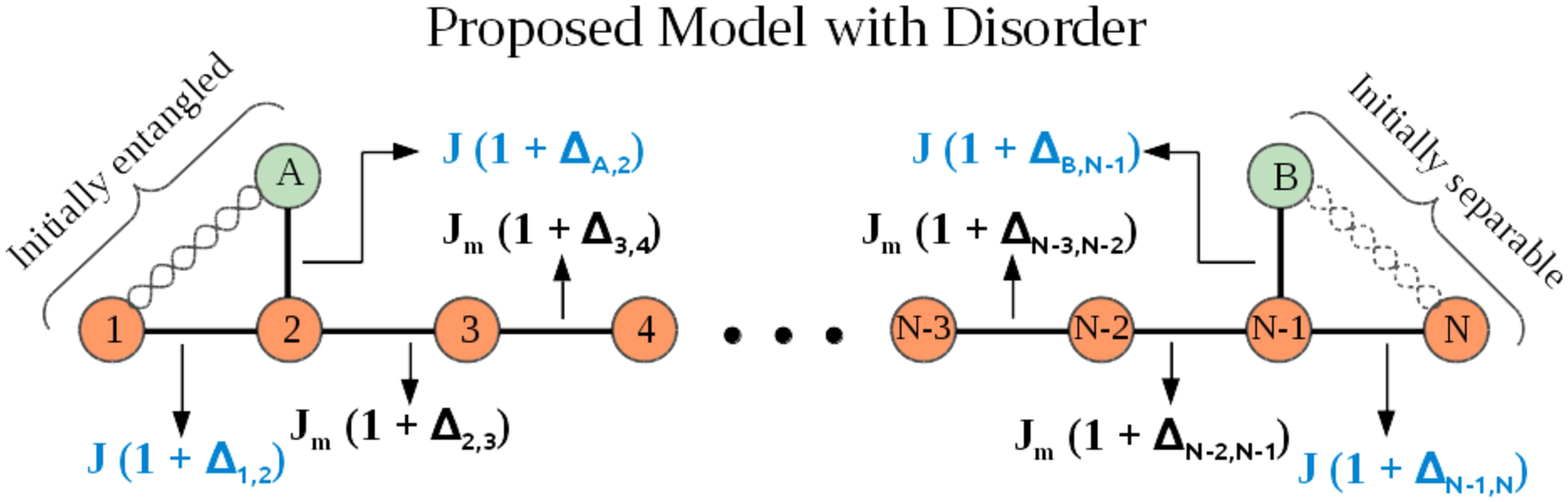}
\caption{\label{fig5}%
Schematic representation of the proposed model
in the presence of static disorder. Each one of the coupling constants 
between two qubits are independently chosen from random uniform 
distributions centered about their optimal values. In other words, coupling constant
$J_{ij}$ between qubits $i$ and $j$ are set at each realization to $J_{ij}(1 + \Delta_{ij})$,
where $\Delta_{ij}$ is a random number chosen from a continuous uniform distribution 
ranging from $-p$ to $p$. We can think of $p$ as
the maximal percentage deviation about the optimal value for $J_{ij}$ in a given disorder realization.
At each realization of disorder in this work,  
the 
coupling constants were assumed to fluctuate within the same percentage 
range $p$.
}
\end{center} \end{figure}

We now want to investigate how the almost perfect pairwise
entanglement transmission capacity of the proposed model is affected by imperfections
in designing the coupling constants among the qubits. Specifically, we want to study
how static disorder affects the dynamics of this system (see Fig. \ref{fig5}).  

\begin{figure}[!ht]
\begin{center}
\includegraphics[width=7cm]{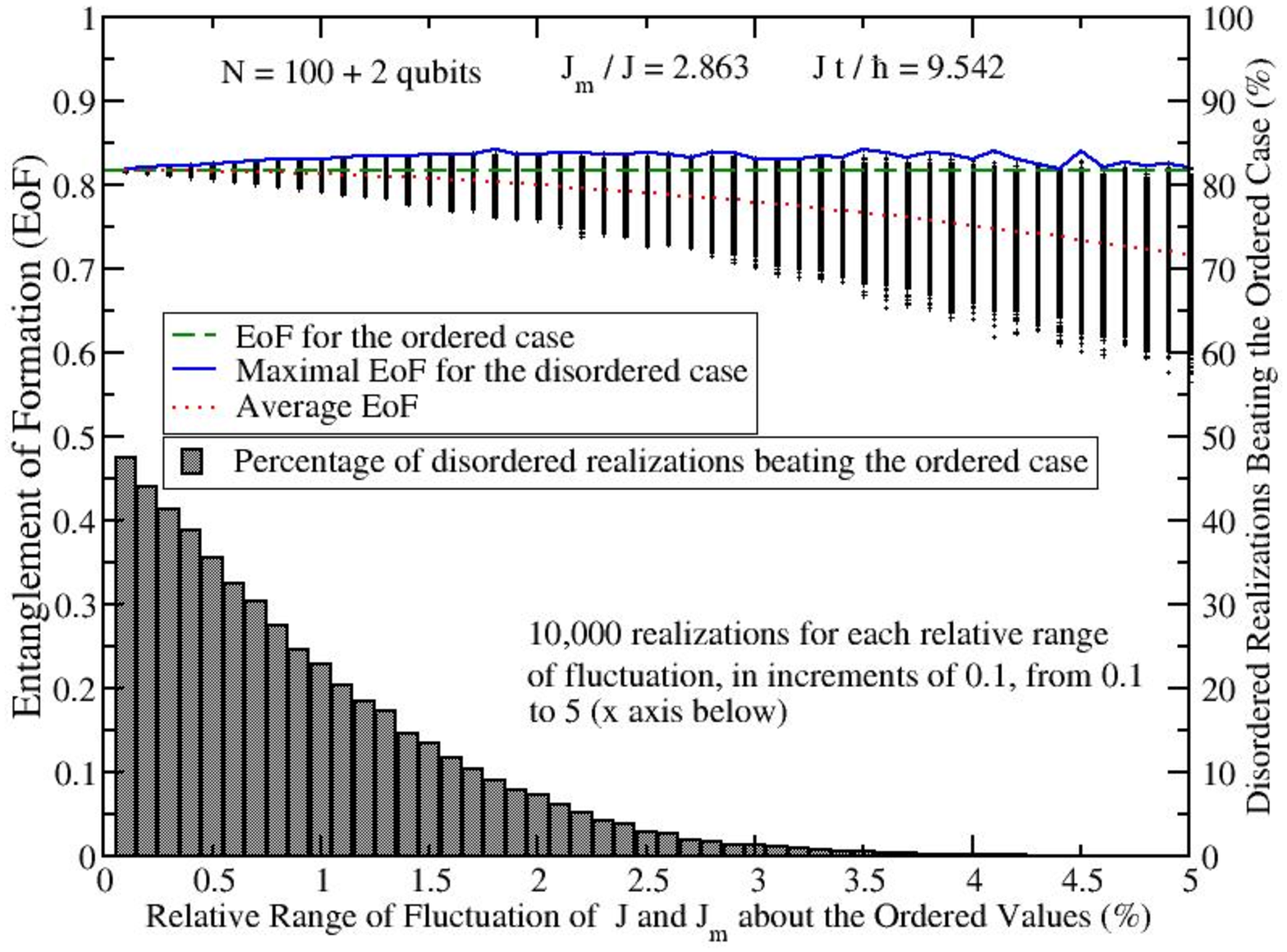}\vspace{.15cm}
\includegraphics[width=7cm]{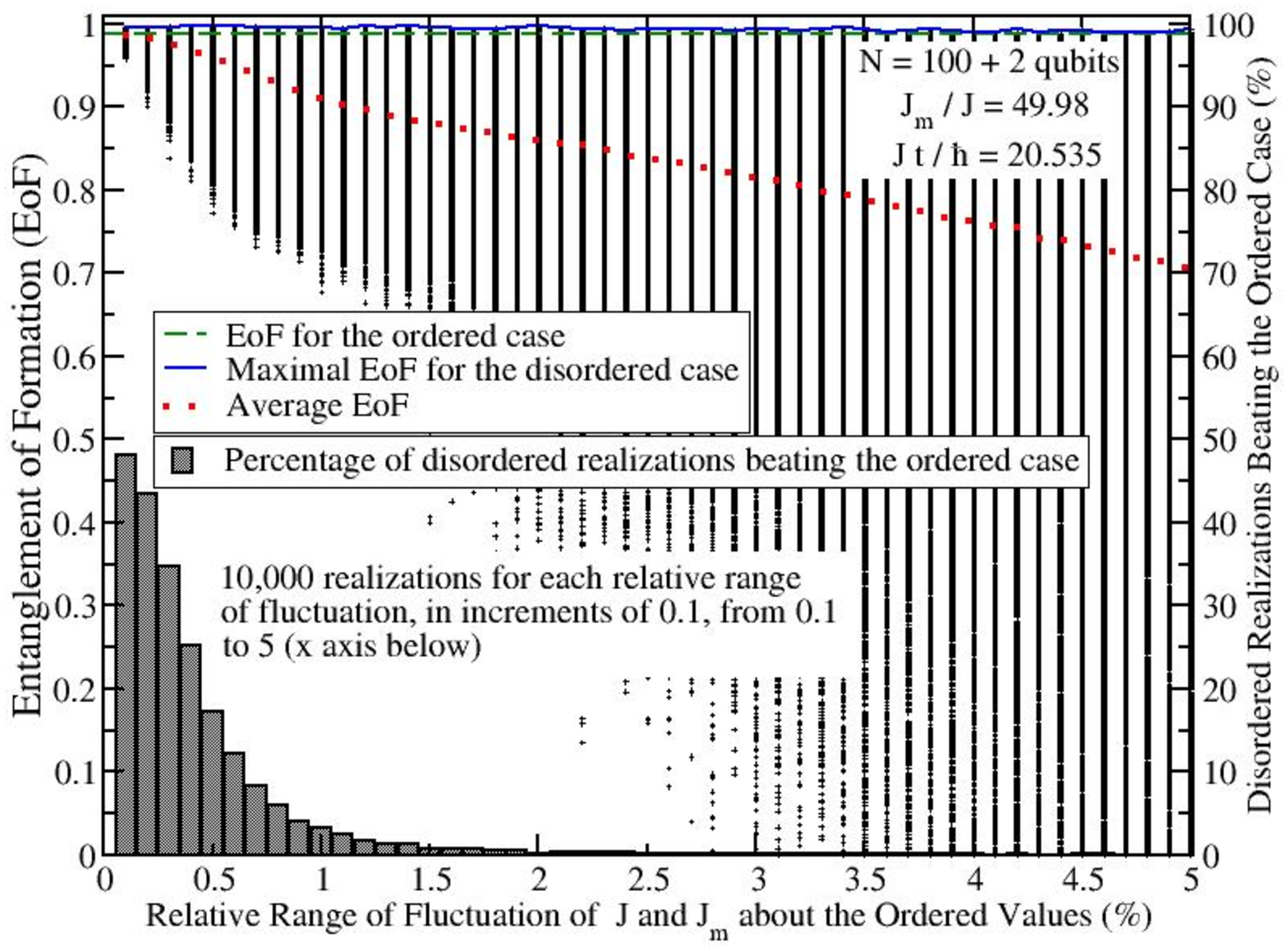}
\caption{\label{fig6e7}%
Upper panel: Here we have the proposed model with $N= 100 + 2$ qubits.
We implemented $10000$ disorder realizations
for each percentage deviation $p$ (x-axis), starting at $p=0.1\%$ and going up to $p=5\%$, in 
increments of $0.1\%$. The data represented by the three curves and by the thousands of small crosses 
for each value of $p$ were computed
at the time $Jt/\hbar = 9.542$. This is the time in which the optimal transmission of entanglement in the clean system 
should occur when we set $J_m/J = 2.863$. This value of $J_m/J$ is the one yielding the best performance
when the search for the optimal coupling constant was restricted from 
$J_m/J =0$ to $5$ (see upper panel of Fig. \ref{fig3e4}).  
The dashed-green curve gives the value of the optimal entanglement 
transmitted to Bob for the clean system. The solid-blue line gives the greatest entanglement 
transmitted out of the $10000$ disorder realizations for each value of $p$. The red-dotted curve gives the average value
of the entanglement transmitted after $10000$ realizations. For every $p$ the entanglement transmitted at each one of the
$10000$ realizations are denoted by the small crosses plotted in the figure. 
The histograms give for every $p$ the percentage 
of the $10000$ disordered realizations outperforming the optimal entanglement transmission 
predicted for the ordered case.
Lower panel: Same as upper panel but now the data 
were computed at the optimal time $Jt/\hbar = 20.535$,  the time where the optimal transmission of entanglement 
in the clean system should occur when we set $J_m/J = 49.98$. This value of $J_m/J$ is the one yielding the best performance
when the search for the optimal coupling constant was made from 
$J_m/J =0$ to $50$ (see upper panel of Fig. \ref{fig3e4}).  
}
\end{center}
\end{figure}
\begin{figure}[!ht]
\begin{center}
\includegraphics[width=7cm]{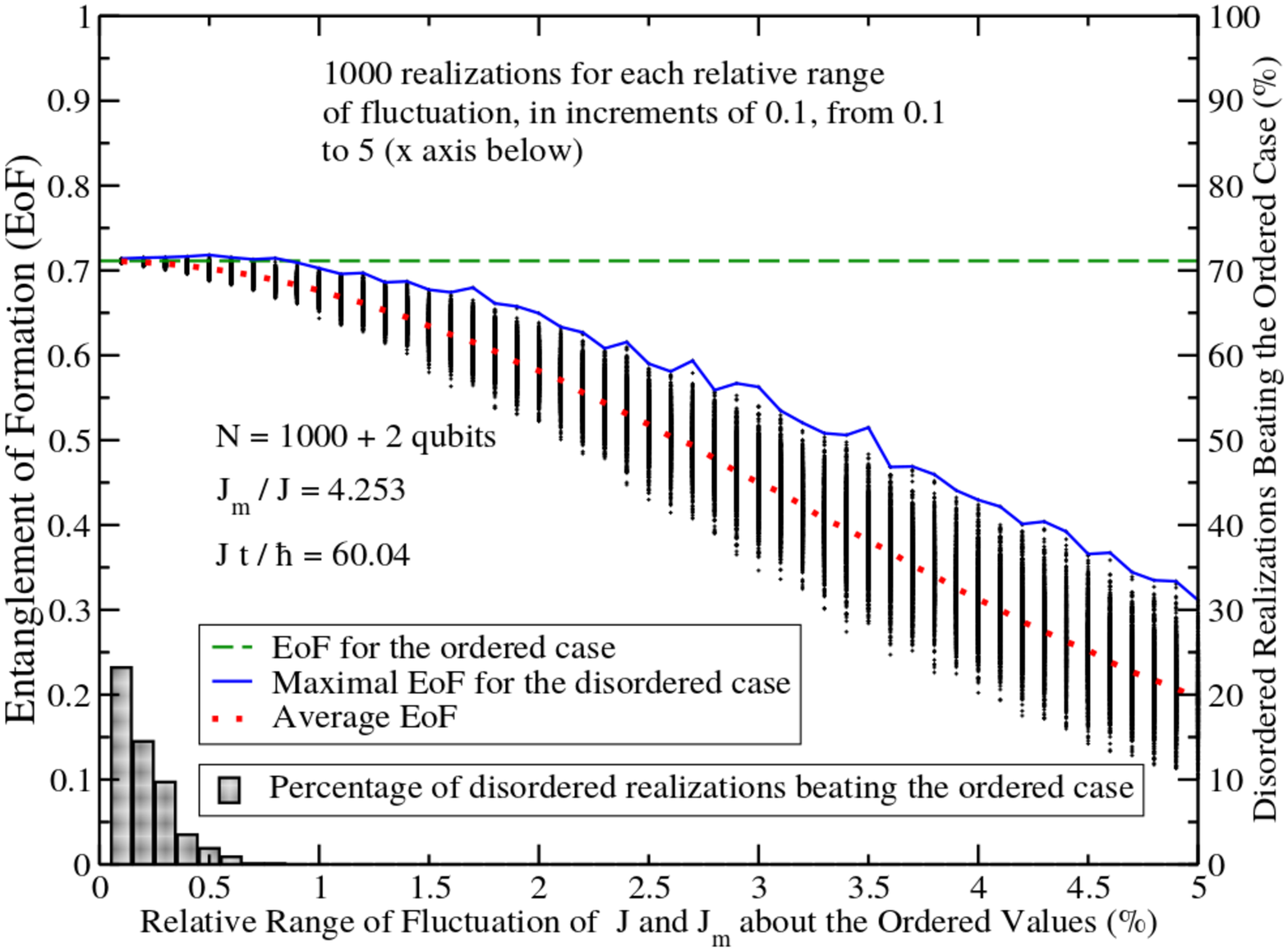}\vspace{.15cm}
\includegraphics[width=7cm]{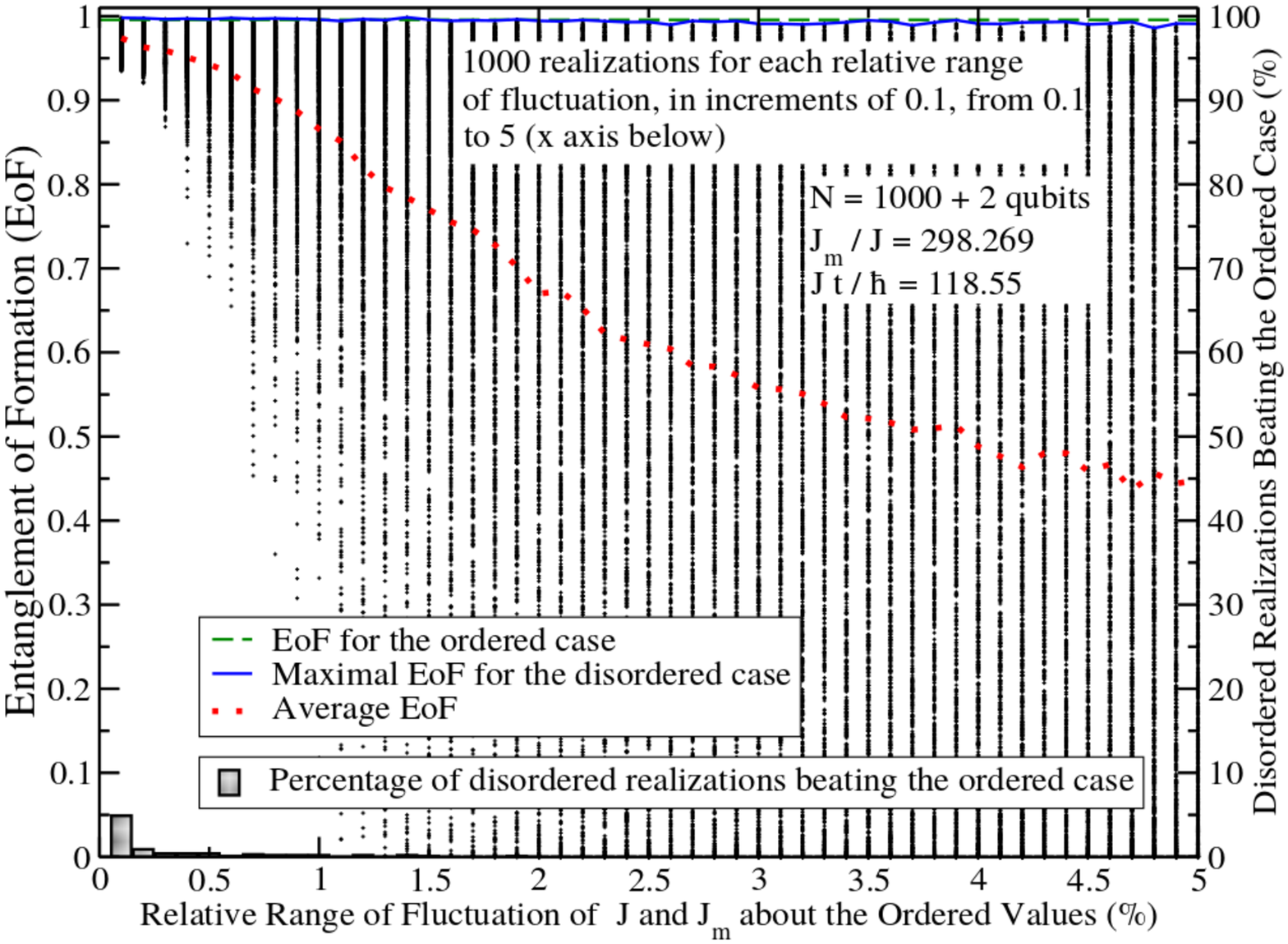}
\caption{\label{fig8e9}%
Upper panel: Here we have the proposed model with $N= 1000 + 2$ qubits. We implemented $1000$ disorder realizations
for each percentage deviation $p$ (x-axis), starting at $p=0.1\%$ and going up to $p=5\%$, in 
increments of $0.1\%$. The data represented by the three curves and by the hundreds of small crosses for each value of $p$ 
were computed
at the time $Jt/\hbar = 60.04$. This is the time in which the optimal transmission of entanglement in the clean system 
should occur when we set $J_m/J = 4.253$. This value of $J_m/J$ is the one yielding the best performance
when the search for the optimal coupling constant was restricted from 
$J_m/J =0$ to $5$ (see lower panel of Fig. \ref{fig3e4}).  
The dashed-green curve gives the value of the optimal entanglement 
transmitted to Bob for the clean system. The solid-blue line gives the greatest entanglement 
transmitted out of the $1000$ disorder realizations for each value of $p$. The red-dotted curve gives the average value
of the entanglement transmitted after $1000$ realizations. For every $p$ the entanglement transmitted at each one of the
$1000$ realizations are denoted by the small crosses plotted in the figure. 
The histograms give for every $p$ the percentage 
of the $1000$ disordered realizations outperforming the optimal entanglement transmission 
predicted for the ordered case.
Lower panel:  Same as upper panel but now the data 
were computed at the optimal time $Jt/\hbar = 118.55$,  the time where the optimal transmission of entanglement 
in the clean system should occur when we set $J_m/J = 298.269$. 
This value of $J_m/J$ is the one yielding the best performance
when the search for the optimal coupling constant was made from 
$J_m/J =0$ to $300$ (see lower panel of Fig. \ref{fig3e4}).  
}
\end{center}
\end{figure}

In Fig. \ref{fig6e7}   
we show the results we obtained for the proposed model in the
presence of disorder for a system of $N=100 + 2$ qubits. In the upper panel of Fig. \ref{fig6e7} 
the analysis was
carried out using the coupling constant $J_m/J=2.863$, the optimal value leading to the best
entanglement transmission for the clean system when the search for the optimal coupling constant was restricted from 
$0$ to $5$. The data in the lower panel of Fig. \ref{fig6e7} 
employed $J_m/J=49.98$, the optimal coupling constant giving
the best entanglement transmission for the clean system 
when the search for the optimal setting was implemented from 
$J_m/J=0$ to $50$.

The first distinctive characteristic that we get by analyzing the data is the fact that 
we obtain average results that outperform the standard model, whose transmission of entanglement
is at best $0.4$ (see upper panel of Fig. \ref{fig3e4}).  
In the proposed model the average entanglement
transmission is always greater than $0.7$, even for fluctuations of $\pm 5\%$ about the optimal settings 
for the ordered system (see the dotted-red curves in Fig. \ref{fig6e7}). 
We have also noted that roughly $50\%$ of the disorder realizations lie above the average
value of transmitted entanglement and the other $50\%$ cases below.

Second, for low disorder, let us say for fluctuations about the optimal settings for the clean system below $2\%$, 
we see that disorder does not affect substantially the entanglement
transmission capacity of the proposed model. Actually, 
in many cases of low disorder we see an increase of the entanglement transmitted from Alice to Bob.
This is more common when we work with the optimal setting $J_m/J=2.863$, where for up to 
$1\%$ of disorder we have at least $1$ of every $4$ disorder realizations yielding a performance
surpassing the clean system (see the histograms of the upper panel of Fig. \ref{fig6e7}). 

In Fig. \ref{fig8e9} 
we show the data obtained for the proposed model in the
presence of disorder for a system composed of $N=1000 + 2$ qubits. In the upper panel of Fig. \ref{fig8e9} 
the calculations were
done using the coupling constant $J_m/J=4.253$, the optimal value furnishing the best
entanglement transmission for the ordered system among all $J_m/J$ lying between $0$ to $5$. 
In the lower panel of Fig. \ref{fig8e9} 
we used $J_m/J=298.269$, the optimal coupling constant that gives 
the best entanglement transmission for the ordered system 
when $J_m/J$ ranges from $0$ to $300$.

Here, and similarly to the case of $100 + 2$ qubits, 
we have a reasonable resilience to disorder, specially in the low disorder scenario. 
Again we have that the average entanglement transmission, being about $0.2$ in the worst case of $5\%$ 
disorder (upper panel of Fig. \ref{fig8e9}), 
is greater than the optimal one for the ordered standard model, which is 
$\approx 0.12$ (lower panel of Fig. \ref{fig3e4}). 
And if we focus on the low disorder sector (below $1\%$ in this case),
we have an average transmitted EoF of about $0.65$ for the case of $J_m/J=4.253$ and
an average EoF of about $0.85$ when $J_m/J=298.269$. 
We also have, but now only for very low disorder, disorder 
realizations leading to a greater performance than the one predicted for the clean system. 
We should also add that the more qubits we add to the system, the more sensitive 
it is to disorder, as a straightforward comparison between
Figs. \ref{fig6e7} and \ref{fig8e9} reveal.
A more detailed 
analysis of the influence of disorder and noise on the model proposed here will be given elsewhere.

\begin{figure}[!ht] \begin{center}
\includegraphics[width=7cm]{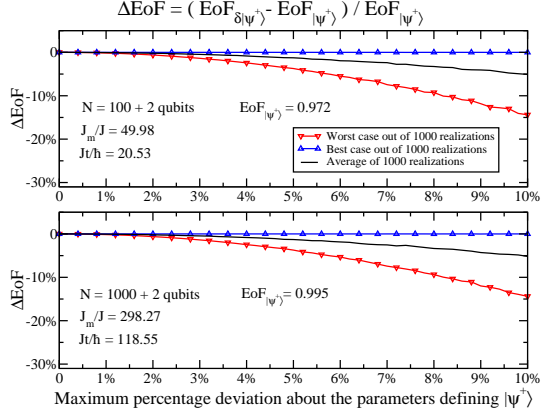}
\caption{\label{fig_extra}%
Upper panel: We work with the proposed model with $N= 100 + 2$ qubits and for each value of $p$ (x-axis) we study the
entanglement transmission efficiency for $1000$ different input states $\delta|\Psi^+\rangle$ as explained in the text. The greater the
value of the x-axis the more the states $\delta|\Psi^+\rangle$ deviate from the Bell state $|\Psi^+\rangle$.
The values of the coupling constants among the qubits $J_m/J$ and the adimensional time $Jt/\hbar$ in which Bob
measures the entanglement content of his qubits are the ones optimizing the entanglement transmission for the
Bell state $|\Psi^+\rangle$ (see Fig. \ref{fig3e4}). $\mbox{EoF}_{|\Psi^+\rangle}$ and  $\mbox{EoF}_{\delta|\Psi^+\rangle}$
are the entanglement measured by Bob at the time $Jt/\hbar$ shown in the figure when the input states are
$|\Psi^+\rangle$ and $|\delta\Psi^+\rangle$, respectively.  
Lower panel: Same as above but now we have $N= 1000 + 2$ qubits.
}
\end{center} \end{figure}

We end this section studying how resilient the present proposal is to imperfections in the preparation of the input
state $|\Psi^+\rangle=(|01\rangle + |10\rangle)/\sqrt{2}$. We assume now that the initial state is given
by $\delta|\Psi^+\rangle= \alpha|01\rangle+\sqrt{1-\alpha^2}e^{-i\gamma}|10\rangle$, where
$\alpha = (1+\Delta\alpha)/\sqrt{2}$ and $\gamma = 2\pi(1+\Delta\gamma)$.
When $\Delta\alpha=\Delta\gamma = 0$ we recover the state $|\Psi^+\rangle$ and we can see
$\Delta\alpha$ and $\Delta\gamma$ as quantifying, respectively, the percentage deviations about the correct
probability amplitude of the state $|01\rangle$ and its relative phase with $|10\rangle$
that give exactly the state $|\Psi^+\rangle$. Note that by studying the evolution of the state $\delta|\Psi^+\rangle$
for several values of $\Delta\alpha$ and $\Delta\gamma$ we are also studying the efficiency of the proposed model
to transmit other states than a Bell state.

In Fig. \ref{fig_extra} we show the results obtained when we allowed
$\Delta\alpha$ and $\Delta\gamma$ to be given by two independent continuous uniform distributions ranging
from $\pm p$. For every value of $p$, with $p$ changing from $0$ to $10\%$ in increments of $0.2\%$, we implemented
$1000$ simulations either using a chain of $100$ qubits (upper panel) or one with $1000$ qubits (lower panel).
Looking at Fig. \ref{fig_extra}
it is clear that for deviations lower than $2\%$ no appreciable change is seen in the entanglement
sent from Alice and received by Bob. Moreover, even when we work with a large deviation of $10\%$, we see an average reduction of
$5\%$ and in the worst scenario
a reduction of only $15\%$ in the entanglement reaching Bob. In this last case, it means that instead of receiving an EoF $\approx 0.97$
Bob gets an EoF $\approx 0.82$, which is still much higher than the EoF transmitted in the optimal scenario using the standard
model and assuming a perfect input state, namely, EoF $\approx 0.40$ for $N=100$ qubits and EoF $\approx 0.12$ for $N=1000$ qubits.

\section{Single excitation transmission}
\label{single}

\begin{figure}[!ht] \begin{center}
\includegraphics[width=8cm]{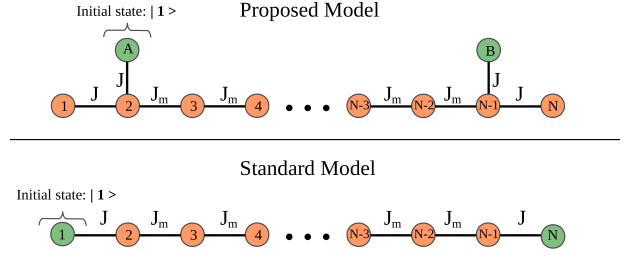}
\caption{\label{fig10}%
Schematic representation of the proposed (upper panel) and
standard (lower panel) models arranged to transmit the single excitation $|1\rangle$ from Alice 
to Bob. The goal here is to transmit the state $|1\rangle$ from qubit $A$ (or $1$) to qubit $B$ (or $N$).
}
\end{center} \end{figure}

For completeness, and also to validate the program we wrote 
to solve the proposed model, we have also studied the transmission
of a single excitation, namely the state $|1\rangle$, from Alice to Bob 
(see Fig. \ref{fig10}). 

\begin{figure}[!ht]
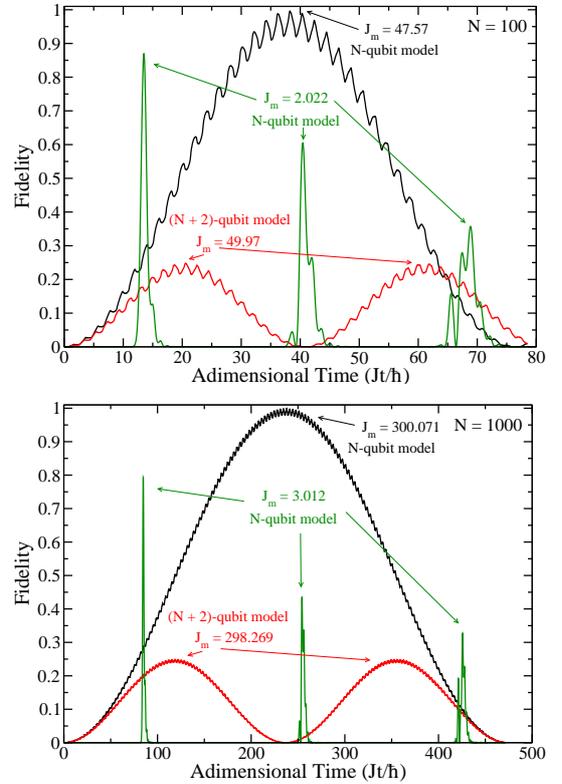

\begin{center}
\includegraphics[width=7cm]{N100Qubits_oneExcitation.eps}\vspace{.15cm}
\includegraphics[width=7cm]{N1000Qubits_oneExcitation.eps}
\caption{\label{fig11e12}%
Upper panel: Single excitation transmission from qubit $A$ ($1$) with Alice
to Bob's qubit $B$ ($N$) for the proposed (standard) model when $N=100$ qubits.
The two curves with the greatest peaks correspond to the standard model ($N$-qubit model) while
the curve with the lowest peaks is related to the proposed model ($N+2$-qubit model). 
Alice's qubit is initially the state $|1\rangle$ while
all the other qubits start at the state $|0\rangle$.  
For simulation purposes, we set $J_{\!_A}=\tilde{J}_{\!_A}=J_{\!_B}=\tilde{J}_{\!_B}=J=1$ for
the proposed model and $J_{\!_A}=J_{\!_B}=J=1$ for the standard model. In both models we evolved the system from
$Jt/\hbar=0$ to $25\pi$ for every value of $J_m$ ranging from $0$ to $50$ in increments of $0.01$,
saving the value of $J_m$ giving the best single excitation transmission (greatest fidelity)
within the time span
given above. Subsequently we moved about this value
in increments of $0.001$ in order to get the optimal
$J_m$ shown in the figure. For the standard model we also give the optimal $J_m$ restricting
its range from $0$ to $5$ (green curve showing the first peak above). 
Lower panel: Same as upper panel but now we have $N=1000$ qubits and we evolved the system from
$Jt/\hbar=0$ to $150\pi$ for every value of $J_m$ ranging initially from $0$ to $300$ in
increments of $0.01$. }
\end{center}
\end{figure}

We first tested our program searching for the optimal configurations for the standard model leading to
the best single excitation transmission. The results we obtained were exactly the ones reported in \cite{woj05}, 
an important validation of our code. Then we turned on the coupling constants $\tilde{J}_A$ and $\tilde{J}_B$, setting
them to $J$, and searched for the optimal arrangements leading to the greater single excitation transmission
for the proposed model.

The approach employed in the search for the optimal settings leading to the most efficient 
single excitation transmission was
similar to that already reported in the entanglement transmission scenario. 
Being more specific, we have evolved the system for every value 
of $J_m/J$ within a predetermined range, changing it in fixed small increments. 
The system always started with Alice's qubit $A$ (or $1$) given by
$|1\rangle$ and the remaining ones
by $|0\rangle$. For every value of $J_m/J$ we have evolved the system up to a given dimensionless 
time $Jt/\hbar$, computing along the way the fidelity
between qubit $B$ and the state $|1\rangle$ (or between $N$ and $|1\rangle$) 
according to the recipe given in Sec. \ref{tools}

In Fig. \ref{fig11e12} 
we report the simulations for the case of $100$ and $1000$ qubits, 
respectively. The same strategy, time span, and coupling constant ranges employed previously in the
entanglement transmission analysis of Sec. \ref{results} were used here. The only exception was for the case of
$1000$ qubits, where the time span employed ranged from $0$ to $Jt/\hbar=150\pi$.

It is clear looking at Fig. \ref{fig11e12} that 
for the transmission of a single excitation the standard model is the
optimal choice, yielding almost perfect single excitation transmission. 
The proposed model, however, is not indicated for this task. Actually, the proposed model can never excel in this task
since the Hamiltonian and the initial condition in this case are symmetric with respect to the exchange between qubits $N$ and $B$. 
This implies that the density matrix describing qubits $N$ and $B$ are identical at any time and we thus must conclude
that we can never have qubit $B$ in the state $|1\rangle$ and qubit $N$ in the state $|0\rangle$.

\section{Conclusion}
\label{conclusion}

In this work we have studied the transmission of a maximally entangled two-qubit state (Bell state) 
from one place (Alice) to another one (Bob) along a graph.
The graph used here
was a slightly modified linear spin chain, where in addition to $N$ qubits arranged along a chain we have
added two extra ones, namely, qubit $A$ with Alice, interacting with qubit $2$, and qubit $B$ with Bob,
interacting with qubit $N-1$ (see Fig. \ref{fig1}). The specific model employed here was
the isotropic XY model (XX model) describing a spin-1/2 chain. 

We have shown that a strictly linear chain described by the XX model cannot accomplish satisfactorily such a task without 
the aid of modulated external magnetic fields or without modulated coupling constants among the qubits. 
This was our main motivation to go beyond a strictly linear chain. Indeed, in the simple scenario of no external fields, 
no modulation, and only local control of the coupling constants at the end points of the chain, a strictly linear chain 
furnishes very poor results in transmitting pairwise entanglement, with a rapid decreasing 
efficiency as we increase the size of the chain. 

Interestingly, nevertheless, by working with the slightly modified linear chain as described above, 
we can transmit in an almost perfect fashion a maximally entangled two-qubit state from one  
end of the chain to the other one. Furthermore, this is possible 
without external fields or modulation of the coupling constants among the qubits. We only
need to match the values of the coupling constants among the end point spins with Alice with those with
Bob to accomplish almost perfect pairwise entanglement transmission for chains
of any size (see Fig. \ref{fig2}).

We have also tested the present proposal against imperfections in its construction, namely, we tested 
how it responds to random variations of the coupling constants about their optimal values. We have shown
that for moderate static disorder we still have a good performance, beating the pairwise entanglement transmission efficiency 
of the ordered strictly linear chain.

Moreover, for small disorder, i.e., up to a fluctuation of $\pm 1\%$ about the optimal settings
for the clean system, we have obtained several realizations of disorder giving better results than those  
predicted by the optimal ordered and noiseless case.

We have also tested the proposed model in its capacity to transmit a single excitation from Alice to Bob,
comparing its efficiency to that of the strictly linear chain. We have shown that in this case the 
strictly linear chain is the best choice.  

Finally, it would be interesting to extend the present work in at least the following two directions. 
First, how would the present model respond to an open dynamics, when the whole chain interacts with its
environment. Second, what would happen if the chain interacts with a thermal reservoir at temperature $T$. 
Is it possible to model and solve this problem in an efficient way? If yes, how could that be done? 
Investigations of the efficiency of the proposed model in those two scenarios and, to be honest, of the many 
models given in previous works cited 
in this manuscript, are important to assert the 
true usefulness of all those proposals in a real world operation of, for example, 
a yet-to-be-built quantum computer based on solid state devices.

\section*{Acknowledgments}
RV thanks CNPq (Brazilian National Council for Scientific and Technological Development)
for funding and GR thanks CNPq and CNPq/FAPERJ (State of Rio de Janeiro Research Foundation) for financial support through the National Institute of
Science and Technology for Quantum Information.

\end{document}